\documentstyle[prb, aps, graphicx]{revtex}
\tighten

\newcommand\braket[1]{\left\langle #1 \right\rangle}

\title{Fractionalized Phase in an $XY-Z_2$ Gauge Model}

\author{R.~D.~Sedgewick\thanks{rds@physics.ucsb.edu},
D.~J.~Scalapino\thanks{djs@physics.ucsb.edu}, and
R.~L.~Sugar\thanks{sugar@physics.ucsb.edu}}
\address{Department of Physics, University of California, Santa Barbara, CA 93106}
\begin{document}
\draft

\date{January 8, 2001}
\maketitle

\begin{abstract}

We study a model with fractional quantum numbers using Monte Carlo
techniques.  The model is composed of bosons interacting though a
$Z_2$ gauge field.  We find that the system has three phases: a phase
in which the bosons are confined, a fractionalized phase in which the
bosons are deconfined, and a phase in which the bosons are condensed.
The deconfined phase has a ``topological'' order due to the degeneracy
in the ground state of the gauge field.  We discuss an experimental
test proposed by Senthil and Fisher that uses the topological order to
determine the existence of a deconfined, fractionalized phase.

\end{abstract}
\section{Introduction}

A theoretical framework for constructing model many-body systems that
can exhibit phases with fractionalized quantum numbers has been
proposed by Senthil and Fisher.\cite{SF99}  Here, we report results
obtained from a numerical simulation of one such model and examine a
recently proposed experimental test for detecting the fractionalized
phase.\cite{results,Aug2000}

\begin{figure}
\begin{center}
\includegraphics{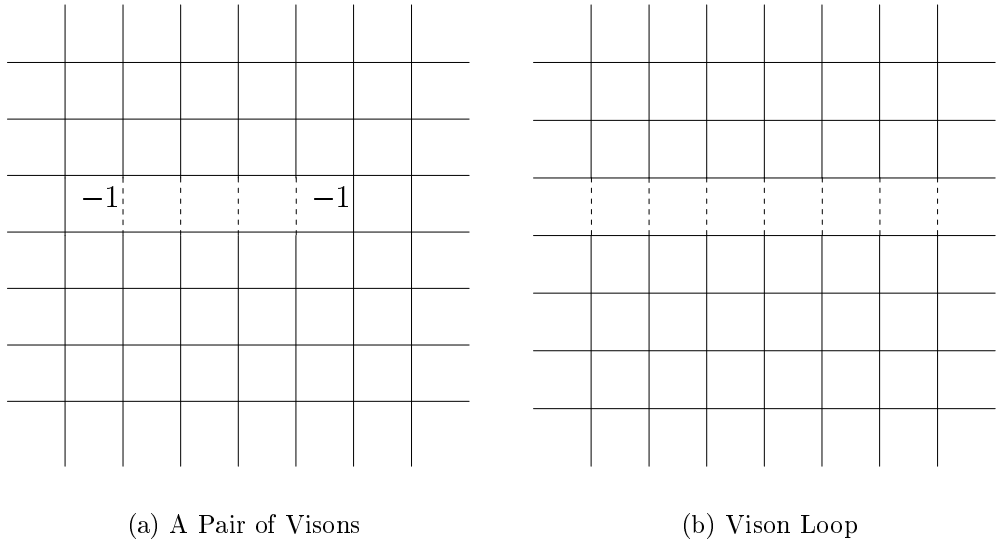}
\caption{Schematic representation of the 2D quantum system.  Solid lines denote $\sigma_{ij}=1$, dotted lines denote $\sigma_{ij}=-1$.  All visons on the lattice are labeled $-1$.}
\label{fig:visons}
\end{center}
\end{figure}

The quantum many-body model that we will study consists of
``chargons'' coupled to a fluctuating $Z_2$ gauge field in two spatial
dimensions.  It was obtained by Senthil and Fisher by considering the
special case of $s$-wave pairing with an even number of electrons
per unit cell, and integrating out the spinon degrees of freedom. 
This model has the advantage of being straightforward to simulate,
while allowing a test of some of the underlying ideas associated
with the fractionalized phases.
The Hamiltonian for the 2D quantum lattice has the form
\begin{equation}
 H = -J \sum_{\langle ij\rangle}\sigma^z_{ij}\left(b^\dagger_i b_j + h.c.\right) +
A \sum_i n^2_i -
K\sum_{\Box} \left[\prod_{\Box} \sigma^z_{ij}\right] - h
\sum_{\langle ij\rangle} \sigma^x_{ij} \label{qham}
\end{equation}
where the chargon creation operator is $b^\dagger_i= e^{i\hat{\phi}_i}$
and $n_i$ is conjugate to $\hat{\phi}_i$ so that $[\hat{\phi}_i, n_j]=i\,
\delta_{ij}$. The gauge field operator on the link between
nearest-neighbor sites $i$ and $j$ is $\sigma^z_{ij}$, and
$\prod\limits_\Box \sigma^z_{ij}$ is the product of gauge operators
around a plaquette. The sum $\langle ij\rangle$ is over nearest neighbor
sites on the 2D spatial lattice. A ``vison'' excitation consists of a
plaquette for which $\prod\limits_\Box \sigma^z_{ij}$ is equal to $-1$.
Visons are always joined in pairs by a string of plaquettes with two
links flipped relative to the rest of the links in the vicinity.  A
schematic example of this is shown in Fig.~\ref{fig:visons}(a). Here
and in the text, we have taken the flipped links to have $\sigma^z_{ij}=-1$
for definiteness.

\begin{figure}
\begin{center}
\includegraphics{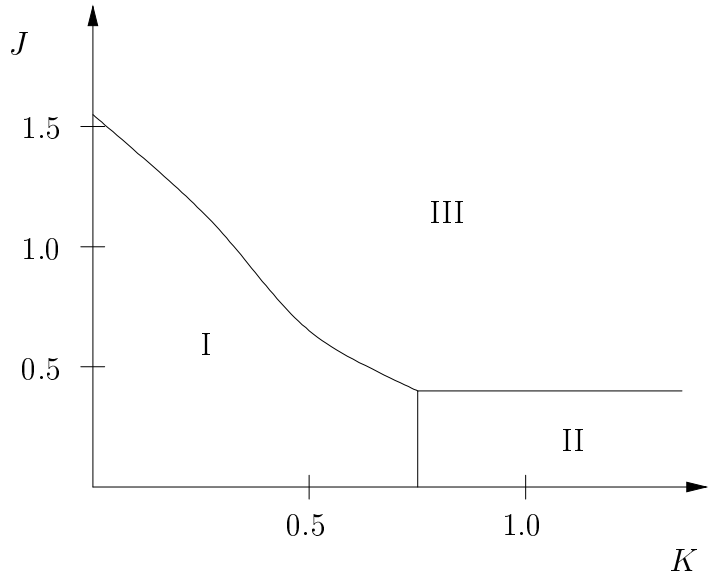}
\end{center}
\caption{Phase diagram showing confined phase (I), deconfined phase (II), and boson condensed phase (III).}
\label{fig:phase}
\end{figure}

For the purpose of carrying out simulations, we work with the 3D (two
space and one Euclidean time) classical action associated with
the Hamiltonian of Eq.~(\ref{qham}),
\begin{equation}
S=-J \sum_{\langle ij\rangle} \sigma_{ij} \cos\left(\phi_i-\phi_j\right)
- K\, \sum_{\Box}\left[\prod_{\Box} \ \sigma_{ij}\right].
\label{oneone}
\end{equation}
 This action has an $XY$ angular variable $\phi_i$ corresponding to
the eigenvalue of the operator $\hat{\phi}_i$ and a $Z_2$ gauge field
$\sigma_{ij}=\pm 1$ corresponding to the eigenvalue of the operator
$\sigma^z_{ij}$. Here, $\langle ij\rangle$ indicates all nearest
neighbor sites on the 3D lattice.  This action has rotational symmetry
as well as a local $Z_2$-gauge symmetry in which $\phi_i$ at a site
transforms to $\phi_i+\pi$ and all of the $\sigma_{ij}$ gauge fields
linked to the $i^{\rm th}$ site change sign.  As discussed in Ref.\
\onlinecite{SF99}, this model is expected to have the type of phase
diagram illustrated in Fig.~\ref{fig:phase}. Both the gauge field
$\sigma_{ij}$ and the $XY$ field $\phi_i$ are disordered in region I
when $J$ and $K$ are small.  In the 2D quantum version, this corresponds to the
fluctuating $\sigma^z_{ij}$ gauge field confining the chargons so that
there are no free $b^\dagger_i$ excitations, only $(b^\dagger_i)^2$
excitations.  In region III, the $XY$ rotation symmetry is broken, as
well as the $Z_2$ gauge symmetry.  This is just the usual $XY$ phase
with a finite helicity modulus. Here, chargon pairs $(b^\dagger_i)^2$
condense to form a superfluid.

Region II corresponds to a ``fractionalized'' (unconfined) phase. We find
that the $\phi$ field is disordered, as in region I, but the $Z_2$
gauge field is ordered, or equivalently in the quantum version, the visons are
gapped.\cite{SU2}  This allows the chargon pairs to ``fractionate,'' and
individual $b^\dagger_i$ chargon excitations are present in the
quantum version.  As one enters the superconducting phase, region III,
it is the $b^\dagger_i$ field that condenses.

The fractionalized phase, region II, is characterized by a
``topological'' order.  That is, on a manifold with a non-trivial
topology, the ground state of the 2D quantum system has a degeneracy
that depends upon the topology.  With periodic boundary conditions the
2D quantum system has the topology of a torus.  A non-trivial
topological excitation occurs if a string of plaquettes with two
flipped links associated with a vison pair cuts through the torus, as
illustrated schematically in Fig.~\ref{fig:visons}(b). In order to
minimize the energy, the $\phi$ field has a discontinuity of $\pi$
(mod $2\pi$) across the flipped links denoted by the dashed lines of
Fig.~\ref{fig:visons}(b). Because the $\phi$ field does not have
long-range order in region II, this disturbance dies out within a
correlation length, and its energy does not grow with the lattice
size.  When a vison loop threads the 3D torus used in our simulations,
this topological configuration is trapped because the free energy
barrier the system must go over to reach the no-vison state grows as
the lattice size, $L$, which goes to infinity in the bulk limit.  As
proposed by Senthil and Fisher, the existence of this topological
order can be probed by driving this system into the superfluid state
III, where the $\phi$ field does have long-range order.  In this case,
when a trapped vison is present in region II and the system is driven
from region II to region III (by, for example, increasing $J$), the
$\pi$ phase shift in $\phi$ associated with the vison will induce a
circulating current of bosons.\cite{results}  Thus, by going into the
superconducting phase III one can look for trapped visons by measuring
the boson current. If they exist, the trapped visons tell us that we
have come from a ``fractionalized'' phase.

Abelian gauge theories have been studied extensively by high-energy physicists
since the earliest days of lattice gauge theory.\cite{KOGUT} The work most
closely related to our own is the study of the Abelian Higgs model with a
$Z_2$ gauge field coupled to an Ising matter
field,\cite{BDI,FradkinShenker,CREUTZ} and a $U(1)$ gauge field coupled to an
$XY$-matter field.\cite{FradkinShenker,EinhornSavit,JKS} Here we study a $Z_2$
gauge field coupled to an $XY$ field. More recently, an action similar to that
of Eq.~\ref{oneone}, but with an $O(3)$ matter field, was used by Lammert,
Rokhsar, and Toner to study a classical model for nematics.\cite{LRT}

The existence of three distinct phases in the model we study is consistent 
with conclusions drawn from the study of more general Abelian Higgs models,
in which the Higgs field is not in the fundamental representation of
the gauge group.\cite{FradkinShenker} In particular, as pointed out
by Senthil and Fisher\cite{SF99}, each of the phases discussed here
has an analogue in the $O(3)$ model of nematics.\cite{LRT}
By contrast, the $Z_2$ gauge theory coupled to an Ising matter
field has only a confined and a deconfined phase with the Higgs and
confined phases being analytically connected,\cite{FradkinShenker,CREUTZ}
as is expected in general for Abelian Higgs theories in which the
matter field is in the fundamental representation of the gauge 
group.\cite{FradkinShenker}

In Section 2, we will discuss Monte Carlo results for the Polyakov loop (the
product of $\sigma_{ij}$ wrapped periodically around the lattice) and the
helicity modulus that give us numerical results for the phase diagram. Then in
Section 3, we will discuss visons and the Senthil-Fisher test for
fractionalization.  Section 4 contains our conclusions.

\section{Phase Diagram\label{sec:phase}}

Using the 3D Euclidean action, we measure the Polyakov loop and the helicity
modulus to find the transitions in both the gauge field and the bosonic field.
On our lattices the Polyakov loop provides a useful probe of the gauge field,
as we will discuss.  The helicity modulus measures the stiffness of the bosonic
field to rotating the spins.  It can indicate the phase of the boson field,
since in our $XY$ spin formulation the superconducting state is characterized
by a finite spin stiffness.

\begin{figure}
\begin{center}
\includegraphics[width=3.5in, angle=-90]{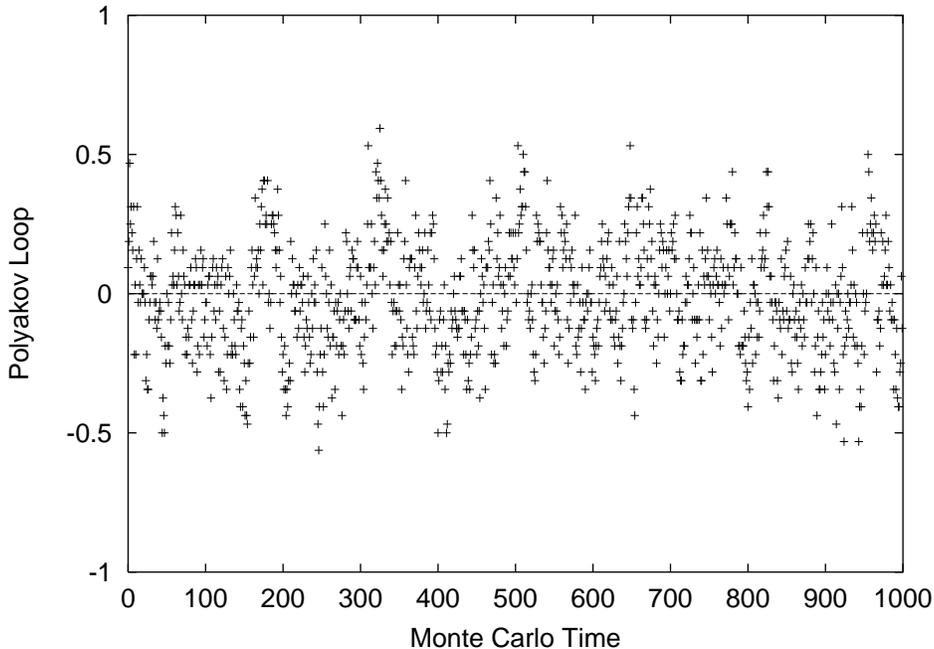}
\caption{The average of the Polyakov loop across the lattice as a function of Monte Carlo time for J=0, K=0.69.  This is done on an $8^3$ lattice using local Metropolis updating.}
\label{fig:polymc1}
\end{center}
\end{figure}

\begin{figure}
\begin{center}
\includegraphics[width=3.5in, angle=-90]{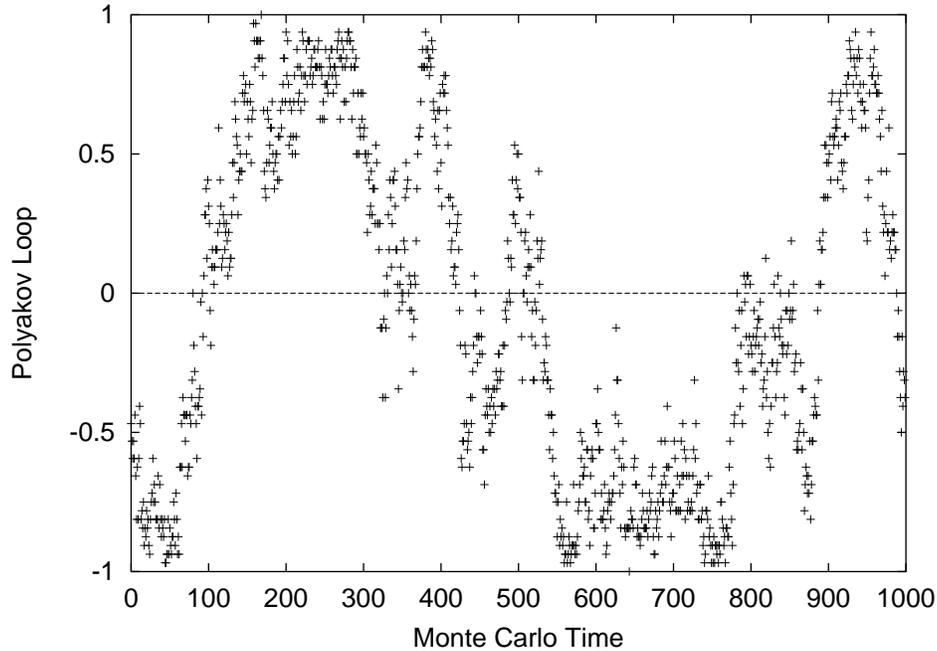}
\caption{The average of the Polyakov loop across the lattice as a function of Monte Carlo time for J=0, K=0.73.  This is done on an $8^3$ lattice using local Metropolis updating.}
\label{fig:polymc2}
\end{center}
\end{figure}

It is well known that for $J=0$ the action of Eq.~\ref{oneone} gives rise to a
second-order phase transition on an infinite three-dimensional
lattice.\cite{KOGUT} The confined (strong coupling) phase is characterized by
area law behavior of the Wilson loops, and the deconfined (weak coupling) phase
by perimeter law behavior. However, on a finite lattice there is a cross-over,
rather than a {\it bona fide} phase transition. For the relatively small
lattices on which we perform our simulations, measurement of the Polyakov loop
provides a convenient way to locate the cross-over.  A Polyakov loop is the
product of a line of $\sigma_{ij}$ wrapped periodically around the lattice,
\begin{equation}
P_{\hat{\mu}}= \sigma_{ij} \sigma_{j\ell}\ldots \sigma_{mn} \sigma_{ni},
\end{equation}
where all the links are pointing in the $\hat{\mu}$ direction.
For strong coupling it vanishes order by order in perturbation theory,
and in our simulations we find that it fluctuates about zero, as is
seen in Fig.~\ref{fig:polymc1}. At
weak coupling an expansion about the state in which
all the $\sigma_{ij}$ have the same sign gives, to leading order,\cite{KOGUT}
\begin{equation}
P=e^{-2Le^{-8K}},
\label{expfalloff}
\end{equation}
where $L$ is the number of lattice points in the temporal direction. A
reference state in which the Polyakov loop has the same magnitude, but opposite
sign, can be obtained by the addition of a vison through the torus that flips
all the links in the temporal direction between two adjacent time slices. By
tracking $P$ in our simulations, we observe the system tunneling between these
two degenerate states, as is seen in Fig.~\ref{fig:polymc2}.  On our finite lattice $P$
averages to zero because of cancellations between states in which it takes on
positive and negative values, but above the transition the tunneling occurs
much less frequently than our period of observation so that we obtain a
non-zero average. It follows from Eq.~\ref{expfalloff} that for $L\to \infty$
the jump in $P$ goes to zero.  However, in our simulations, $L=8$, and the
cross-over is in the neighborhood of $K=0.7$. For these values the magnitude of
the Polyakov loop at the crossover is approximately $0.94$.

\begin{figure}
\begin{center}
\includegraphics[width=3.5in, angle=-90]{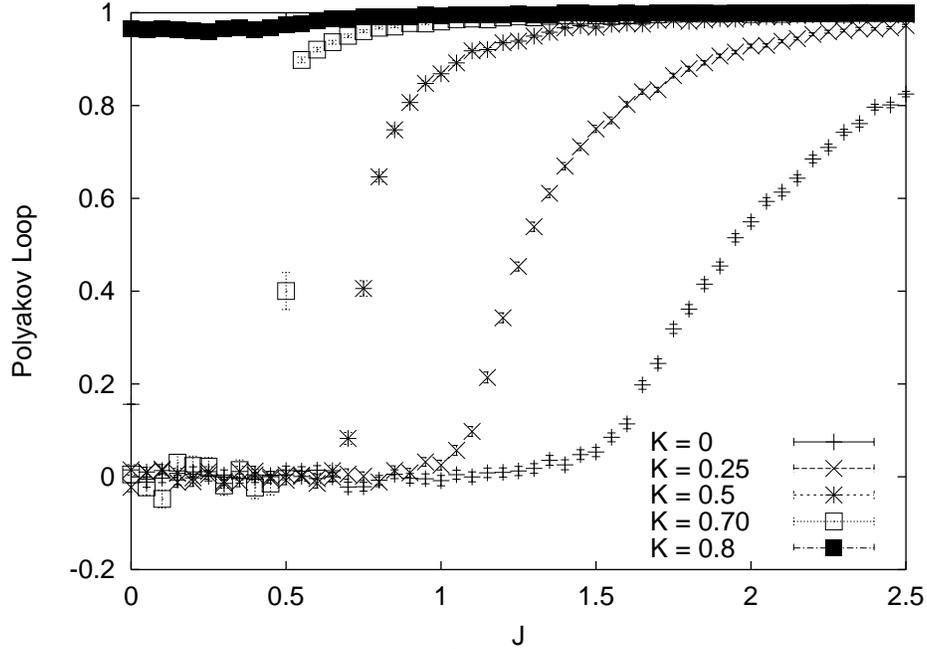}
\caption{Expectation value of the Polyakov loop on an $8^3$ lattice along lines of constant $K$. We took 150 measurements at each point with each measurement separated by 35 Monte Carlo steps.  The system is allowed to equilibrate for 200 Monte Carlo updates between points.  }
\label{fig:poly}
\end{center}
\end{figure}

\begin{figure}
\begin{center}
\includegraphics[width=3.5in, angle=-90]{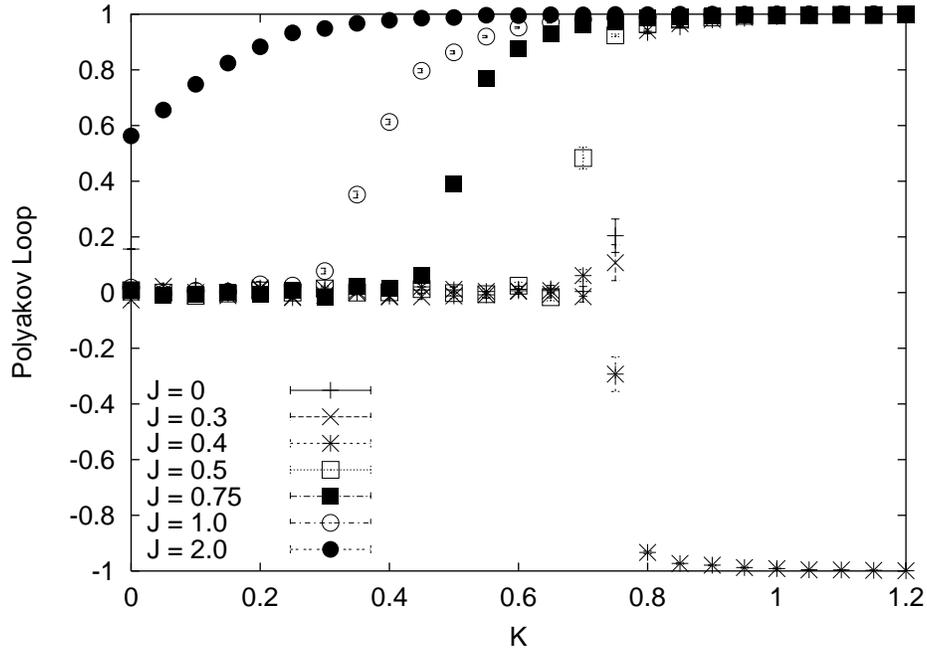}
\caption{Expectation value of the Polyakov loop on an $8^3$ lattice along lines of constant $J$.
We took 150 measurements at each point with each measurement separated by 35 Monte Carlo steps.  The system is allowed to equilibrate for 500 Monte Carlo updates between points.
A vison is trapped in the run where $J=0.4$.}
\label{fig:polyj}
\end{center}
\end{figure}

We then measured the Polyakov loop along lines of constant $J$ and $K$.  Figure
\ref{fig:poly} shows the expectation value of Polyakov loops along lines of
constant $K$.  These Monte Carlo measurements were taken using a local
Metropolis updating scheme on a lattice with $8^3$ sites.  We took 150
measurements for each point with each measurement separated by 35 Monte Carlo
updates of the lattice.  After the measurements at a point are done, $J$ is
increased and the system is allowed to equilibrate for 200 Monte Carlo updates.
In this case, each run is started with the system completely ordered so that no
vison loops get frozen into the system.  Figure \ref{fig:polyj} shows the
expectation value of Polyakov loops along lines of constant $J$.  The
measurements were taken in the same way with $K$ increased during the run
instead of $J$.  Additionally, 500 Monte Carlo steps were used in between
measurements to equilibrate.  For this figure each run is started from a
completely disordered state.  When the system enters the deconfined phase (II)
from the confined phase (I) it has the opportunity to trap a vison. A trapped
vison changes the expectation value of the Polyakov loop from $1$ to $-1$ and accounts for the run with $J=0.3$.

\begin{figure}
\begin{center}
\includegraphics[width=3.5in, angle=-90]{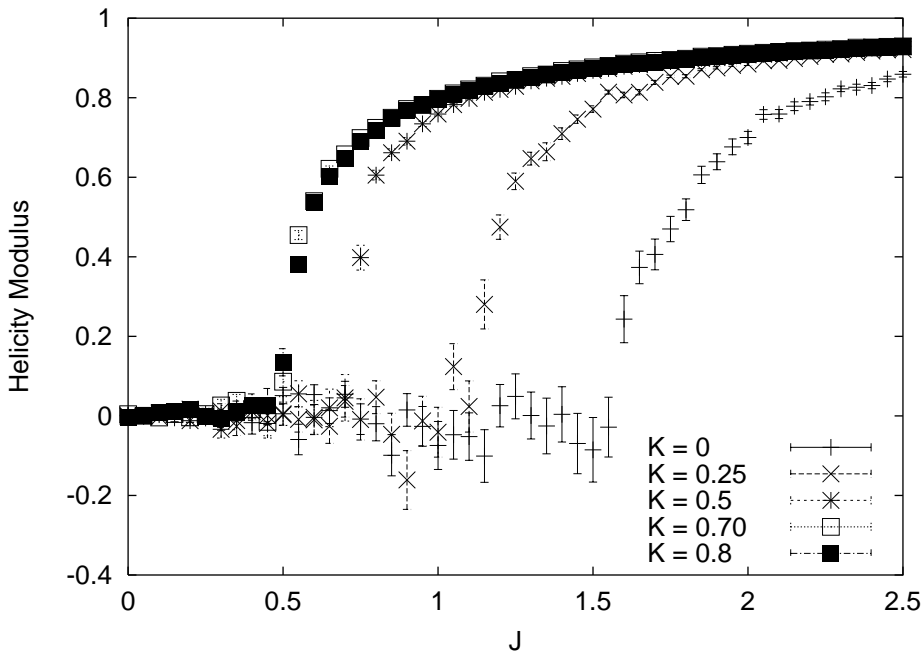}
\caption{Expectation value of the helicity modulus on a $8^3$ lattice along lines of constant $K$. We took 150 measurements at each point with each measurement separated by 35 Monte Carlo steps.  The system is allowed to equilibrate for 200 Monte Carlo updates between points.  }
\label{fig:helmod}
\end{center}
\end{figure}

The standard form of the helicity modulus for the $XY$ model is not
invariant under the $Z_2$ gauge transformation. It can be made
gauge invariant by inserting factors of the gauge field.  This gives a helicity
modulus of the form
\begin{equation}
\Upsilon_{\hat{\mu}}/J = \frac{1}{N} \braket{\sum_{\braket{ij}}\sigma_{ij}
\cos(\phi_i-\phi_j)(\hat{\epsilon}_{ij}\cdot\hat{\mu})^2}
-\frac{J}{N} \braket{\left(\sum_{\braket{ij}} \sigma_{ij}
\sin(\phi_i-\phi_j)\hat{\epsilon}_{ij}\cdot\hat{\mu}\right)^2},
\end{equation}
where $\hat{\mu}$ points along the bonds in the $\hat{x}$, $\hat{y}$,
or $\hat{z}$ directions and $\hat{\epsilon}_{ij}$ is a unit vector
pointing from the $i^{\rm th}$ lattice site to the $j^{\rm th}$
lattice site.  Since the helicity modulus measures the stiffness of
the spins, it is $0$ where the bosons are disordered and finite where
the bosons have long-range order.\cite{XYmodel}  The helicity modulus
measured along lines of constant $K$ is shown in
Fig.~\ref{fig:helmod}.  These measurements were taken in the same
way as the measurements of the Polyakov loop: 150 measurements at each
point separated by 35 Monte Carlo steps between measurements and 200
Monte Carlo measurements between points.  The transition is rounded
due to finite size effects.

The results found by looking at the expectation value of the helicity modulus
and the Polyakov loop were used to construct the phase diagram shown in
Fig.~\ref{fig:phase}.  The transitions were taken to be when the observable is
statistically non-zero.  The nature of the transition between the deconfined
phase and the confined phase and between the deconfined phase and the condensed
phase are understood from studying the Ising model and the XY model, and the
boundary between the confining phase and the condensed phase has been studied
by Senthil and Fisher.\cite{quest}  This is a transition at which both the
gauge field and the boson field order.  This phase transition occurs down to
$K=0$ where there is no plaquette term in the action, implying that it is the
ordering of the bosons that forces the gauge field to order.

\section{Visons\label{sec:visons}}

As discussed in the introduction, the fractionalized phase II has a topological
order caused by the presence of visons threading the torus.  These visons are
trapped, as previously noted.  Senthil and Fisher recently proposed a method
for detecting trapped visons by driving the system between the fractionalized
phase II and phase III, in which the $\phi$ field is
condensed.\cite{results,Aug2000}  In our numerical experiments the presence of
visons can be determined by observables such as the Polyakov loop that measure
the gauge field directly, while in real-world systems this is typically not
possible.  It is therefore necessary to be able to determine the topological
order without observing the gauge field directly.  However, as discussed by
Senthil and Fisher, the boson field can be used to probe the topology on the
lattice in the condensed phase, in which it has long-range order.
Anti-periodic boundary conditions generated by a vison threading the torus
cause the bosons to gradually twist by $\pm \pi$ from one side of the vison
string to the other. This gives rise to a non-zero gauge invariant current,
\begin{equation}
I_{\hat{\mu}} =J \braket{\sum_{i} \sigma_{i\, i+\hat{\mu}}\sin(\phi_i-\phi_{i+\hat{\mu}})},
\end{equation}
which can be measured.  Here $\hat{\mu}$ is a spatial unit vector perpendicular
to the vison string.  In this way, measuring the gauge-invariant current can
determine the topological order of the lattice.  Thus, when a single vison
string threads the torus we expect a boson current to flow corresponding to the
$\pi$ change in the phase produced by the vison.

In the ordered phase, the fluctuations in the boson phase lead
to a renormalization of J,
\begin{equation}
J_r = J \braket{\sigma_{i\,i+\hat{\mu}} \cos(\phi_i-\phi_{i+\hat{\mu}})}.
\end{equation}
As the bare coupling $J$ increases and one goes deep into the boson
condensed phase, the phase fluctuations decrease and $J_r/J$ goes to
$1$.  In a real physical measurement, the quantity entering the
circulation or flux is the renormalized coupling $J_r$, so that a
measurement of $I_{\hat{\mu}}$ gives $J_r \pi$. For a finite $L^3$
lattice we expect that when a vison is trapped and the system is
switched into the condensed boson phase III by increasing $J$, one
will find
\begin{equation}
I_{\hat{\mu}}/J_r = \pm L\sin(\frac{\pi}{L}),
\end{equation}
which goes to $\pm\pi$ as $L$ goes to infinity.

\begin{figure}
\begin{center}
\includegraphics[width=3.5in, angle=-90]{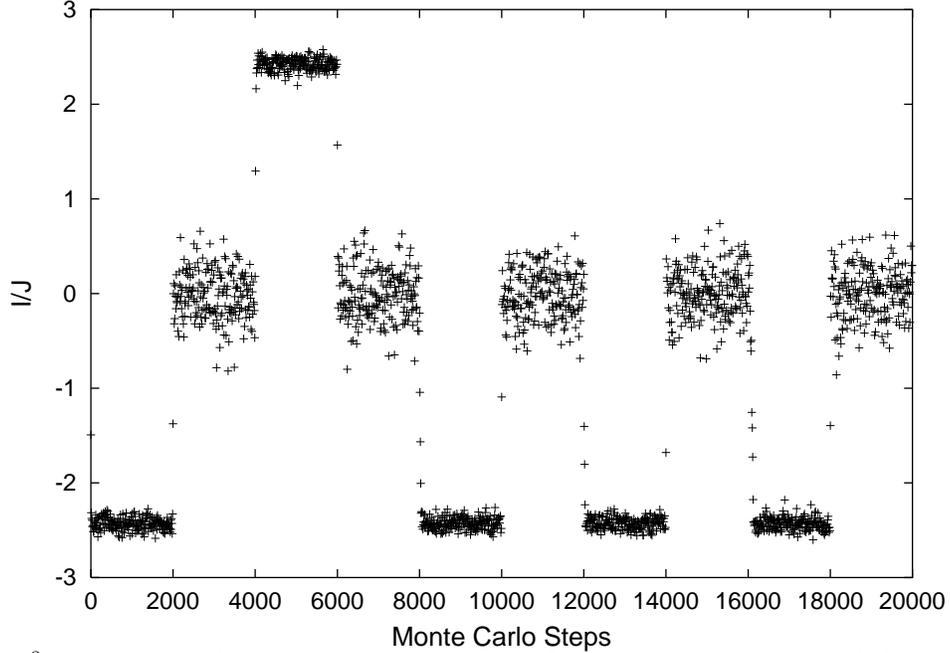}
\caption{The $8^3$ lattice is moved between the boson condensed phase ($K=1$, $J=1$) and the deconfined phase ($K=1$, $J=0.25$) every 2000 Monte Carlo steps.  Note that $I/J$ is not $\pm\pi$ in the boson condensed phase due to fluctuations of the bosons and finite size effects as discussed in the text.}
\label{fig:visa}
\end{center}
\end{figure}

The lattice is initially prepared at $K=1$ and $J=1$ in the condensed
phase with a single vison string threading the torus and a gradual
twist of the bosons by $\pi$.  In a real-world system, this would be
achieved by threading an $h c/2 e$ magnetic flux quantum through the
sample.\cite{results,Aug2000}  If the system is then moved to the
fractionalized phase by decreasing $J$ to $0.25$, the boson current
disappears, but the vison remains trapped so that the boson current
returns with the same magnitude if the system is moved back to the
condensed phase.  We have done this numerically on an $8^3$ lattice
using our Euclidean action, with the results shown in
Fig.~\ref{fig:visa}. Here we have plotted $I_{\hat{\mu}}/J$ and we see
that this ratio is non-zero in the condensed phase, signifying the
presence of a vison that remains trapped in the fractionalized
phase. For $K=1$ and $J=1$ on an $8^3$, lattice $J_r/J$ is measured to
be $0.81$ so that
\begin{equation}
\frac{I_{\hat{\mu}}}{J}=\pm\frac{J_r}{J} 8 \sin\left( \frac{\pi}{8}\right) \simeq \pm 2.48,
\end{equation}
in agreement with Fig.~\ref{fig:visa}.  Note that the boson current
can be either positive or negative reflecting the direction of the
gradual twist induced by the vison.

\begin{figure}
\begin{center}
\includegraphics[width=3.5in, angle=-90]{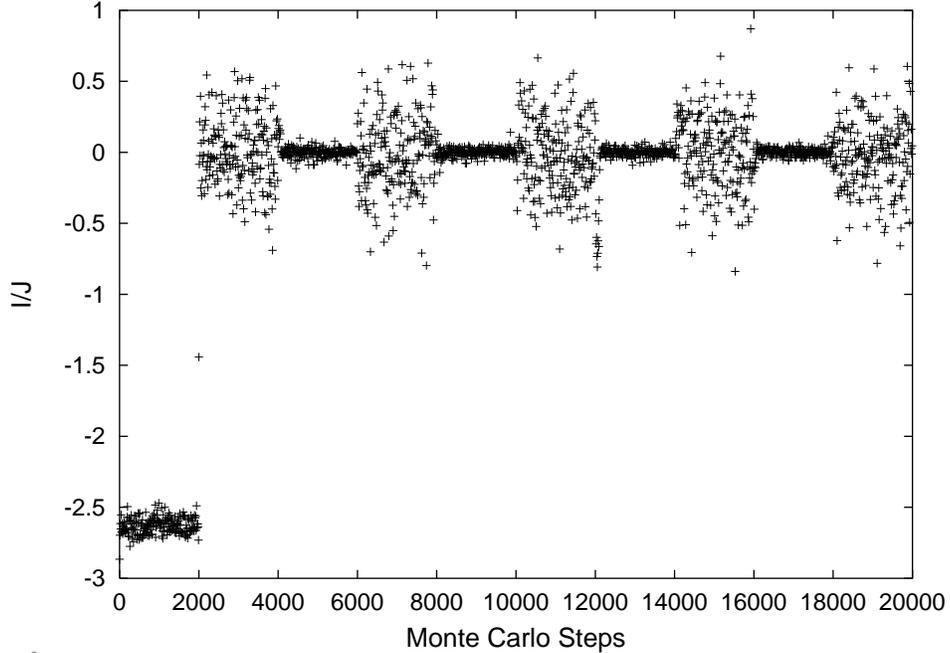}
\caption{The $8^3$ lattice is moved between the boson condensed phase ($K=0.5$, $J=1.5$) and the confined phase ($K=0.5$, $J=0.25$) every 2000 Monte Carlo steps.}
\label{fig:visb}
\end{center}
\end{figure}

If the system with a single trapped vison is moved to a phase without
fractionalization, the confined phase I, the vison can escape.  For
example, at $K=0.5$ and $J=1.5$ we have prepared the system in the
condensed phase with a single vison string threading the torus and a
boson current, as shown over the first $2000$ sweeps in
Fig.~\ref{fig:visb}.  For $K=0.5$ and $J=1.5$ on an $8^3$ lattice
$J_r/J$ is measured to be $0.88$, so that initially $I_{\hat{\mu}}/J
\simeq 2.69$. Then the system is taken into the confined phase I by
decreasing $J$ to $0.25$ at constant $K=0.5$.  When, after another
$2000$ Monte Carlo steps, the system is taken back into the boson
condensed phase III by increasing $J$ to $1.5$, the boson current is
seen to vanish.  This means that the vison that was initially
trapped in the condensed phase escaped when $J$ was decreased to
$0.25$ with $K=0.5$.  The boson current remains zero through further
cyclings between the phases.  This measurement shows that
$(J=0.25, K=0.5)$ corresponds to a point in the confined phase.

\begin{figure}
\begin{center}
\includegraphics[width=3.5in, angle=-90]{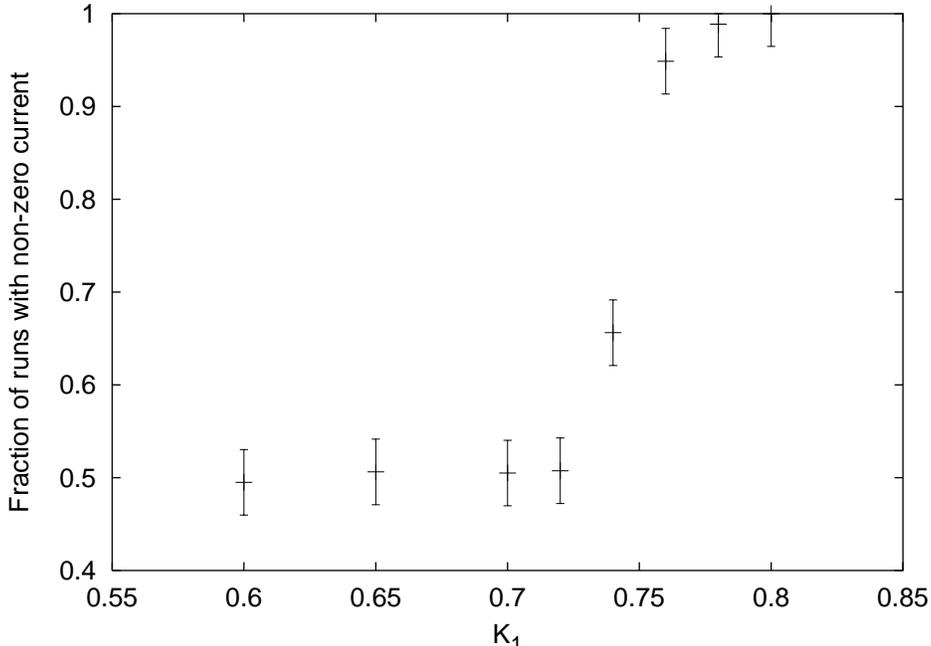}
\caption{Fraction of $800$ measurements in which an odd number of visons were detected after a system initially prepared in the fractionalized phase with a vison is heated at a constant rate to $K_1$ and then cooled back to the fractionalized phase.  These results are along the line $J_1=0.2$.  The existence of an odd number of visons is determined by moving the system to the condensed phase and observing the average bosonic current.}
\label{fig:vescape}
\end{center}
\end{figure}

This type of measurement can also be used to determine the line separating the
confined and fractionalized phases.  We can determine whether a given point in
the phase diagram $(K=K_1, J=J_1)$, is in the fractionalized or confined phase
by the following set of measurements.  We start by equilibrating a trapped
vison at the point $(K=1, J=J_1)$, which we know to be in the fractionalized
phase for sufficiently small $J_1$.  We next decrease $K$ to $K_1$, equilibrate
the system, and then increase $K$ back to $1$, where we again equilibrate the
system.  Finally we move to $(K=1, J=1)$, a point at which we know the bosons
are condensed.  If $(K_1, J_1)$ is in the confined phase, then the trapped
vison will escape; however, the system may trap visons when re-entering the
fractionalized phase.  If an odd number of visons are trapped, there will be a
non-zero bosonic current in the condensed phase, while if an even number of
vison are trapped, no bosonic current will flow in the condensed phase.  Thus
we expect to find a current in $50\%$ of the runs. On the other hand, if $(K_1,
J_1)$ is in the fractionalized phase, then the initial vison will remain
trapped, and a non-zero current will be observed in the condensed phase.
Figure~\ref{fig:vescape} shows the fraction of $800$ trials in which we
observed a trapped vison for $J_1=0.2$ and at different values of $K_1$.  The
time to equilibrate the lattice in the condensed phase was highly variable
because of false energy minima created by trapping several visons.  The system
has probability $1/2$ of trapping an odd number of visons in moving from the
confined phase to the fractionalized phase, but when the system is kept in the
fractionalized phase the initial vison remains trapped.  The transition is
rounded due to visons tunneling out of the finite torus close to the
transition.

\section{Conclusion}

In this paper we have reported on an investigation of the XY model
coupled to a $Z_2$ gauge field using Monte Carlo techniques.  This
model was proposed by Senthil and Fisher for the study of
fractionalization.  By observing the Polyakov loop and the helicity
modulus, we have determined the structure of the phase diagram of the
theory.  In addition, we considered an experiment proposed by Senthil
and Fisher to determine the existence of a fractionalized phase.
Using phase III, in which the bosons are condensed, to measure the
existence of a vison, we see that the vison remains trapped in the
fractionalized phase.  A vison that
was trapped in the phase III region is able to escape when the system
is taken into the confined phase I.


\acknowledgements

We would like to thank M.P.A.~Fisher and T.~Senthil for many insightful
discussions. We would also like to thank R.~Moessner, S.~L.~Sondhi and
J.~Zannan for their helpful comments regarding the Polyakov loop.  This work
was supported by Department of Energy grant 85-ER-45197, National Science
Foundation grant CDA96-01954 and by Silicon Graphics Inc.


\end{document}